\documentclass[journal]{IEEEtran}
\usepackage{epsfig,color,amsmath,cite}
\usepackage{amsthm}
\usepackage{amsmath}    
\IEEEoverridecommandlockouts
\usepackage{bm}
\usepackage{epstopdf}
\usepackage{amssymb}
\usepackage{url}
\usepackage{enumitem}
\usepackage{multirow}
\usepackage{hhline}
\usepackage{booktabs}
\usepackage{threeparttable}
\usepackage{algorithm,algorithmic}	
\usepackage{comment}
\setlist[itemize]{leftmargin=*}

\DeclareMathOperator*{\minimize}{minimize}

\DeclareMathOperator*{\subjectto}{subject\ to}

\makeatother
\DeclareMathAlphabet\mathbfcal{OMS}{cmsy}{b}{n}

\newtheorem{theorem}{Theorem}
\newtheorem{mydef}{Definition}



\usepackage{stackengine}

\newcommand{\mat}[1]{\boldsymbol{#1}}

\newcommand{\bmat}[1]{\begin{bmatrix} #1 \end{bmatrix}}

\providecommand{\mA}{\ensuremath{\mat{A}}}

\providecommand{\mC}{\ensuremath{\mat{C}}}

\providecommand{\mI}{\ensuremath{\mat{I}}}

\providecommand{\mL}{\ensuremath{\mat{L}}}

\providecommand{\mO}{\ensuremath{\mat{O}}}
\providecommand{\mP}{\ensuremath{\mat{P}}}

\providecommand{\mY}{\ensuremath{\mat{Y}}}
\providecommand{\mZ}{\ensuremath{\mat{Z}}}

\newcommand{\m}{\boldsymbol}
\allowdisplaybreaks[4]
\usepackage[colorlinks = true,
linkcolor = blue,
urlcolor  = blue,
citecolor = blue,
anchorcolor = blue]{hyperref}

\newcommand{\mbb}[1]{\mathbb{#1}}

\usepackage[framemethod=TikZ]{mdframed}
\mdfdefinestyle{MyFrame}{%
	linecolor=black,
	outerlinewidth=1.25pt,
	roundcorner=1.25pt,
	innerrightmargin=5pt,
	innerleftmargin=5pt,}
	
\usepackage[noabbrev]{cleveref}

\usepackage{mathtools}

\DeclarePairedDelimiter\abs{\lvert}{\rvert}%
\DeclarePairedDelimiter\norm{\lVert}{\rVert}%

\makeatletter
\let\oldabs\abs
\def\abs{\@ifstar{\oldabs}{\oldabs*}}
\let\oldnorm\norm
\def\norm{\@ifstar{\oldnorm}{\oldnorm*}}
\makeatother

\usepackage[english]{babel}
\usepackage[utf8]{inputenc}
\usepackage[super]{nth}

\usepackage{graphicx}
\usepackage{float}
\usepackage[caption = false]{subfig}

\newcommand{\linf}{\mathcal{L}_{\infty}}

\title{\vspace{0.7cm} \LARGE \textbf{Asymmetric Cell Transmission Model-Based, Ramp-Connected Robust Traffic Density Estimation under Bounded Disturbances}}

\author{Suyash C. Vishno$\text{i}^{\dagger}$, Sebastian A. Nugroh$\text{o}^{\ddagger}$, Ahmad F. Tah$\text{a}^{\ddagger}$, Christian Claude$\text{l}^{\dagger}$, and Taposh Banerje$\text{e}^{\ddagger}$
	\thanks{
		$^\dagger$Department of Civil, Architectural, and Environmental Engineering, The University of Texas at Austin, 301 E. Dean Keeton St. Stop C1700, Austin, TX 78712.
		$^{\ddagger}$Department of Electrical and Computer Engineering, The University of Texas at San Antonio, 1 UTSA Circle, San Antonio, TX 78249.
		Emails: scvishnoi@utexas.edu, sebastian.nugroho@my.utsa.edu, ahmad.taha@utsa.edu, christian.claudel@utexas.edu, taposh.banerjee@utsa.edu.
		This work was partially supported by the National Science Foundation under Grants 1636154, 1728629, 1917164, and 1917056.}
}

\begin{document}

\maketitle

\setlength{\abovedisplayskip}{3.5pt}
\setlength{\belowdisplayskip}{3.5pt}
\setlength{\abovedisplayshortskip}{3.1pt}
\setlength{\belowdisplayshortskip}{3.1pt}

\newdimen\origiwspc%
\newdimen\origiwstr%
\origiwspc=\fontdimen2\font
\origiwstr=\fontdimen3\font

\fontdimen2\font=0.65ex

\begin{abstract}
In modern transportation systems, traffic congestion is inevitable. To minimize the loss caused by congestion, various control strategies have been developed most of which rely on observing real-time traffic conditions. As vintage traffic sensors are limited, traffic density estimation is very helpful for gaining network-wide observability. This paper deals with this problem by first, presenting a traffic model for stretched highway having multiple ramps built based on asymmetric cell transmission model (ACTM). Second, based on the assumption that the encompassed nonlinearity of the ACTM is Lipschitz, a robust dynamic observer framework for performing traffic density estimation is proposed. Numerical test results show that the observer yields a sufficient performance in estimating traffic densities having noisy measurements, while being computationally faster the Unscented Kalman Filter in performing real-time estimation.
\end{abstract}

\begin{IEEEkeywords}
	Traffic state estimation, asymmetric cell transmission model, Lipschitz continuous, robust dynamic state estimation, robust $\mathcal{L}_{\infty}$ observer.
\end{IEEEkeywords}


\section{Introduction}
To perform real-time state-estimation and control of traffic networks, robust mathematical models of traffic flow are desirable that can efficiently describe real world traffic phenomena. For this purpose, physics-based traffic flow models from traffic engineering literature are popular. These models can be broadly classified into two classes: \textit{macroscopic models}~\cite{lebacque2005first}, which compute the evolution of traffic density, and \textit{microscopic models}~\cite{li2017vehicle}, which model the interaction between individual vehicles. Macroscopic flow models are more suited for traffic state estimation as they can be easily scaled to large networks due to their lower computational cost which, unlike microscopic models, is independent of the number of vehicles in the network.

Here in this paper, we focus on the state estimation problem for traffic density on a stretched highway by considering the classical macroscopic \textit{Lighthill-Whitham-Richards} (LWR) flow model~\cite{Lighthill1995b,Richards1956}, which is a first order hyperbolic conservation law. In this model, the relationship between the traffic-flux and traffic-density is encoded in the form of a diagram that is called the \textit{fundamental diagram}. 
In this work we utilize the triangular fundamental diagram which uses distinct linear equations
to describe the flow-density relationship in the free-flow and
congestion regimes of traffic flow. 

The cell transmission model (CTM)~\cite{Daganzo1994,Daganzo19952} is a first-order Godunov~\cite{Godunov1959} approximation of the LWR partial differential equation used to simulate the evolution of traffic on roadways and networks. The asymmetric cell transmission model (ACTM)~\cite{Gomes2004,Gomes2006,Gomes2008} is a variant of the CTM that departs from the CTM in its treatment of asymmetric merge junctions such as the on-ramp-highway junctions. Unlike the CTM, it assumes separate allocations of the available space on the highway for traffic from each merging branch, which allows for simpler flow conservation equations at such merges.

The ACTM finds its place in the traffic state estimation literature in few studies pertaining to density estimation on freeways. In \cite{MovingHorizon1} and \cite{MovingHorizon2}, for instance, the authors propose a moving-horizon approach for sensor-fault-tolerant traffic state estimation using the ACTM. Besides these, there are several other studies related to traffic state estimation using other variants of the CTM, references to which can be found in \cite{seo2017traffic}. The above studies utilize a queue model for the on-ramps which does not take into account the density of those ramps, and hence the state of those ramps. In this paper, in order to also estimate the state of the ramps along with the highway, we use density conservation equations to represent the ramp dynamics in a similar manner as the highway sections. In addition to this, these studies use a mode switching scheme to deal with the piecewise linear structure of the model. While such linearized models offer simplicity in contrast with nonlinear dynamic models, they are representative of the dynamics only when the traffic density lies in the vicinity of that point. Consequently, this paper seeks to explicitly study the nonlinear nature of traffic flow model for density estimation under uncertainty. 

To that end, herein we introduce a control-theoretic method to address the traffic density estimation problem on stretched highways connected to multiple ramps under constrained and noisy measurements. The paper contributions and organizations are summarized as follows. In Section \ref{sec:model}, we formulate the dynamics of traffic model on highways connected to on- and off-ramps using ACTM and modified them accordingly so that the densities on both highways and ramps are all covered. 
Next in Section \ref{sec:observer}, we present a robust observer framework developed using the concept of $\mathcal{L}_{\infty}$ stability for discrete-time Lipschitz nonlinear systems. 
This observer extends our prior work \cite{Nugroho2018TrafficJournal}, which presents a similar observer for continuous-time Lipschitz nonlinear systems. 
In Section \ref{sec:numerical}, we demonstrate the performance of our observer for performing density estimation on a simple highway and compare it with \textit{Unscented Kalman Filter} (UKF). Finally, the paper is summarized in Section \ref{sec:summary}.
%
%
%


\noindent {\textbf{Paper's Notation:}} \; Notations $\mathbb{R}^n$ and $\mathbb{R}^{p\times q}$ denote the set of real-valued row vectors with size of $n$ and $p$-by-$q$ real matrices, while $\mathbb{S}^{m}$ denotes the set of symmetric matrices with the dimension of $m$. Specifically, $\mathbb{S}^{m}_{+}$ and $\mathbb{S}^{m}_{++}$ denotes the set of positive semi-definite and positive definite matrices. For any vector $z \in \mathbb{R}^{n}$, $\Vert z\Vert_2$ denotes its Euclidean norm, i.e. 
$\Vert z\Vert_2 = \sqrt{z^{\top}z} $, where $z^{\top}$ is the transpose of $z$. 
For simplicity, the notation $'\ast'$ denotes terms induced by symmetry in symmetric block matrices. Table \ref{tab:notation} provides the nomenclature utilized in the ensuing sections.

\section{ACTM-Based Ramp-Connected Highway Traffic Dynamics Model\vspace{-0.0cm}}\label{sec:model}


\begin{table}[t!]
	\footnotesize	\renewcommand{\arraystretch}{1.3}
	\caption{Paper nomenclature: parameter, variable, and set definitions.}
	\label{tab:notation}
	\centering
	\begin{tabular}{||l|l||}
		\hline
		\textbf{Notation} & \textbf{Description}\\
		\hline
		\hline
		\hspace{-0.1cm}$\Omega$ & \hspace{-0.1cm}the set of highway sections on the stretched highway \\
		\hspace{-0.1cm} & \hspace{-0.1cm}$\Omega = \{ 1,2,\hdots,N \}$ , $N \triangleq \abs{\Omega}$ \\
		\hline
		\hspace{-0.1cm}$\Omega_I$ & \hspace{-0.1cm}the set of highway sections with on-ramps \\
		\hspace{-0.1cm}	&  \hspace{-0.1cm}$\Omega_I = \{ 1,2,\hdots,N_I \}$ , $N_I \triangleq \abs{\Omega_I}$ \\
		\hline
		\hspace{-0.1cm}$\Omega_O$ & \hspace{-0.1cm}the set of highway sections with off-ramps \\
		\hspace{-0.1cm}& \hspace{-0.1cm}$\Omega_O = \{ 1,2,\hdots,N_O \}$, $N_O \triangleq \abs{\Omega_O}$ \\ 
		\hline
		\hspace{-0.1cm}$\hat{\Omega}$ & \hspace{-0.1cm}the set of on-ramps, $\hat{\Omega} = \{ 1,2,\hdots,N_I \}$ , $N_I = |\hat{\Omega}|$\\
		\hline
		\hspace{-0.1cm}$\check{\Omega}$ & \hspace{-0.1cm}the set of off-ramps, $\check{\Omega} = \{ 1,2,\hdots,N_O \}$ , $N_O = |\check{\Omega}| $\\
		\hline
		\hspace{-0.1cm}$T$ & \hspace{-0.1cm}duration of each time step\\
		\hline		
		\hspace{-0.1cm}$l$ & \hspace{-0.1cm}length of each section, on-ramp, and off-ramp\\
		\hline	
		
		\hspace{-0.1cm}$\rho_i [k]$ & \hspace{-0.1cm}traffic density on section $i \in \Omega$ at time $kT$, $k \in \mathbb{N}$ \\
		\hline
    	\hspace{-0.1cm}$q_i[k]$ & \hspace{-0.1cm}traffic flow from section $i \in \Omega $ into the next section\\ 
		\hline
    	\hspace{-0.1cm}$\delta_i[k], \sigma_i[k]$ & \hspace{-0.1cm}demand and supply functions for section $i \in \Omega$\\
		\hline
		
		\hspace{-0.1cm}$\hat{\rho}_i[k]$ & \hspace{-0.1cm}traffic density on the on-ramp of section $i\in \Omega_I $ \\
		\hline
    	\hspace{-0.1cm}$r_i[k]$ & \hspace{-0.1cm}traffic flow into section $i \in \Omega_I $ from the on-ramp\\
		\hline
    	\hspace{-0.1cm}$\hat{r}_i[k]$ & \hspace{-0.1cm}traffic flow into the on-ramp of section $i \in \Omega_I $\\
		\hline
    	\hspace{-0.1cm}$\hat{\delta}_i[k], \hat{\sigma}_i[k]$ & \hspace{-0.1cm}demand and supply functions for the on-ramp  $i \in \Omega_I$ \\ 
		\hline		
		
		\hspace{-0.1cm}$\check{\rho}_i[k]$ & \hspace{-0.1cm}traffic density on the off-ramp of section $i\in \Omega_O$ \\
		\hline
    	\hspace{-0.1cm}$s_i[k]$ & \hspace{-0.1cm}traffic flow from section $i \in \Omega_O $ into the off-ramp\\
		\hline
    	\hspace{-0.1cm}$\check{\delta}_i[k], \check{\sigma}_i[k]$ & \hspace{-0.1cm}demand and supply functions for the off-ramp  $i \in \Omega_O$ \\
		\hline

    	\hspace{-0.1cm}$\check{s}_i[k]$ & \hspace{-0.1cm}traffic flow from the off-ramp of section $i \in \Omega_O $\\
		\hline
		
    	\hspace{-0.1cm}$f_{in}[k]$ & \hspace{-0.1cm}traffic wanting to enter section 1 of the highway\\
		\hline
    	\hspace{-0.1cm}$f_{out}[k]$ & \hspace{-0.1cm}traffic that can leave section $N$ of the highway\\
		\hline
    	\hspace{-0.1cm}$\hat{f}_i[k]$ & \hspace{-0.1cm}traffic wanting to enter the the on-ramp of section $i \in \Omega_I $\\
		\hline
    	\hspace{-0.1cm}$\check{f}_i[k]$ & \hspace{-0.1cm}traffic that can leave the off-ramp of section $i \in \Omega_O $\\
		\hline

    	\hspace{-0.1cm}$\beta_i[k]$ & \hspace{-0.1cm}split ratio for the off-ramp of section $i \in \Omega_O$,\\
    	\hspace{-0.1cm}	&  \hspace{-0.1cm}where $\beta_i[k] \in [0,1]$\\
		\hline
    	\hspace{-0.1cm}$\xi_i[k]$ & \hspace{-0.1cm}occupancy parameter for the on-ramp of section $i \in \Omega_I$\\
    	\hspace{-0.1cm}	&  \hspace{-0.1cm}where $\xi_i[k] \in [0,w_c]$\\
    	\hline
               \hspace{-0.1cm}$\psi(\rho)$ & \hspace{-0.1cm}function that maps traffic density into traffic flow \\
		\hline
		\hspace{-0.1cm}$v_f$ & \hspace{-0.1cm}free-flow speed \\
		\hline
		\hspace{-0.1cm}$w_c$ & \hspace{-0.1cm}congestion wave speed \\
		\hline
		\hspace{-0.1cm}$\rho_m$ & \hspace{-0.1cm}maximum density  \\
		\hline
		\hspace{-0.1cm}$\rho_c$ & \hspace{-0.1cm}critical density \\
		\hline
	\end{tabular}
	\vspace{-0.4cm}
\end{table}

This section presents the dynamic, discrete-time modeling of traffic dynamics on a stretched highway with arbitrary number and location of ramps. 

In this paper, we utilize the \textit{Lighthill-Whitman-Richards} (LWR) Model ~\cite{Lighthill1995b,Richards1956} for traffic flow which is expressed by a partial differential equation given as
\begin{align}
 \frac{\partial \rho (t,d)}{\partial t} + \frac{\partial q(t,d)}{\partial d} = 0 , \label{eq:LWRmodel}
\end{align}
where $t$ and $d$ denote the time and distance; $\rho(t,d)$ denotes the traffic density (vehicles/distance) and $q(t,d)$ denotes the traffic flux (vehicles/time). 
The characteristics of traffic networks model depend on the function that defines the relation between $\rho(t,d)$ and $q(t,d)$, which can be obtained from observing the dynamic behavior of real traffic networks on highways. For this purpose, let $\psi(\cdot)$ be a function that maps traffic density into traffic flow. Such a function is commonly referred to as the \textit{Fundamental Diagram} (FD) and in general has a concave shape and bounded domain. 
Here, we consider a triangular-shaped FD constructed as
\begin{align}
\small\hspace{-0.2cm}\psi\left(\rho(t,d)\right) \hspace{-0.05cm}= \hspace{-0.05cm}
\begin{cases}\hspace{-0.05cm}
v_f\rho(t,d), &\;\mathrm{if} \; 0 \leq \rho(t,d) \leq \rho_c \\
\hspace{-0.05cm}w_c\left(\rho_m-\rho(t,d)\right),&\; \mathrm{if} \;\rho_c \leq \rho(t,d) \leq \rho_m,\\
\end{cases}\label{eq:triangular_model}
\end{align}
From the above relationship, $\psi(\rho(t,d))$ satisfies $\psi(0) = \psi(\rho_m) = 0$. To ensure continuity, we assume that $\psi(\rho(t,d))$ must also satisfy $v_f\rho_c = w_c\left(\rho_m-\rho_c\right)$; see \cite{Gomes2008}.

 To represent the traffic dynamics as a series of difference, state-space equations---a useful bookkeeping for the ensuing discussions---we discretize the LWR Model \eqref{eq:LWRmodel} with respect to both space and time (this is also referred to as the Godunov discretization). This approach allows the highway of length $L$ to be divided into sections (cells) of equal length $l$ and the traffic networks model to be represented by discrete-time equations. 
 
 To ensure computation stability, the Courant-Friedrichs-Lewy condition (CFL) given as ${v_f T}l^{-1}\leq 1$ has to be satisfied. Since each section is of the same length $l$, then we have $\rho(t,d) = \rho(kT,l)$, where $k\in\mathbb{N}$ represents the discrete-time index. For simplicity of notation, from now on $\rho(kT,l)$ will be simply written as $\rho[k]$.\\
\indent The discrete-time flow conservation equation based on ACTM~\cite{Gomes2004,Gomes2006,Gomes2008} can be written as
\begin{align}\label{eq:CTM_flow_conservation}
\hspace{-0.2cm}\rho_i[k+1] &=\rho_i[k]+\frac{T}{l}\big(q_{i-1}[k]+r_{i}[k]-q_{i}[k]-s_{i}[k]\big). 
\end{align}
for $0<i\leq N$, where $N$ is the number of sections that the highway is divided into. To simplify the ensuing expressions, we define two new functions referred to as the demand function $\delta_i[\cdot]$ and the supply function $\sigma_i[\cdot]$. The demand function equals the traffic flux leaving section $i$ through the highway assuming that the next section has infinite storage. The supply function equals the traffic flux that can enter section $i$ through the highway assuming that the previous section has infinite storage. These functions are constructed from the triangular FD given in \eqref{eq:triangular_model}.
The demand function $\delta_i[\cdot]$, with and without off-ramp, can be written as
	\begin{align}\label{eq:demand functions}
	\delta_{i}[k] \hspace{-0.05cm}= \hspace{-0.05cm}\begin{cases}
	\hspace{-0.05cm}\min \big(\bar{\beta_{i}}[k]v_{f}\rho_{i}[k],\bar{\beta_{i}}[k]v_{f}\rho_{c},\frac{\bar{\beta_{i}}[k]}{\beta_{i}[k]}\check{\sigma}_i[k]\big), \;\mathrm{if}\; i\in\Omega_O\\
	\hspace{-0.05cm}\min \big(v_{f}\rho_{i}[k],v_{f}\rho_c\big), \qquad\qquad\qquad\;\;\,\;\mathrm{if}\; i\in\Omega\setminus\Omega_O.
	\end{cases}
	\end{align}
Here $\check{\sigma}_i[k]$ can be given as
\begin{align}
    \check{\sigma}_i[k]=\min\big(w_c(\rho_m - \check{\rho_i}[k]),v_f\rho_c\big).
\end{align} 
In \eqref{eq:demand functions}, the split-ratio $\beta_i[k]$ relates the traffic flow from section $i$ into its off-ramp with the traffic flow from section $i$ into the next section such that
\begin{align}\label{eq:off-ramp relation}
    s_i[k]&=\beta_i[k](s_i[k]+q_i[k]) \implies
  s_i[k]=\frac{\beta_i[k]}{\bar{\beta_i}[k]}q_i[k].
\end{align}
where we define $\bar{\beta_i}[k]\overset{\Delta}{=}1-\beta_i[k]$ to simplify the equations. Similarly, the supply function $\sigma_i[k]$, with and without on-ramp, can be represented as
    \begin{align}\label{eq:supply functions}
	\sigma_{i}[k] \hspace{-0.05cm}=\hspace{-0.05cm} \begin{cases} \hspace{-0.05cm}\min\big(w_{c}(\rho_{m}\hspace{-0.05cm}-\hspace{-0.05cm}\rho_{i}[k])\hspace{-0.05cm}-\hspace{-0.05cm}r_{i}[k],v_f\rho_c\hspace{-0.05cm}-\hspace{-0.05cm}r_i[k]\big),\,\mathrm{if}\, i\in\Omega_I \\
 \hspace{-0.05cm}\min\big(w_{c}(\rho_{m}\hspace{-0.05cm}-\hspace{-0.05cm}\rho_{i}[k]),v_f\rho_c),\,\qquad\qquad\mathrm{if}\, i\in\Omega\setminus\Omega_I,
	\end{cases}
	\end{align}
where $r_i[k]$ is described by the following equation
\begin{align}
    r_i[k]=\min\big(v_f\hat{\rho_i}[k],\xi_i(\rho_m-\rho_i[k]),\frac{\xi_i}{w_c}v_f\rho_c\big).\label{eq:on-ramp flow}
\end{align}
Here $v_f\hat{\rho_i}[k]$ is the demand of the on-ramp in free-flow, and $\frac{\xi_i[k]}{w_c}$ is the fraction of the flow that section $i$ can receive from its on-ramp. Note that (\ref{eq:on-ramp flow}) is a deviation from the original ACTM which uses a queue model for the on-ramps.
We can write the upstream and downstream flows for section $i$ through the highway respectively as
\begin{subequations}\label{eq:CTM_flow_characteristics}\begin{align}
q_{i-1}[k] &= \min \big( \delta_{i-1}[k],\sigma_i[k]\big) \label{eq:CTM_flow_characteristics_1} \\
q_{i}[k] &= \min\big(\delta_i[k],\sigma_{i+1}[k]\big), \label{eq:CTM_flow_characteristics_2}
\end{align}
\end{subequations}
The flow conservation equations for the on-ramp and off-ramp can be written as
\begin{subequations}\label{eq:Flow_conservation_ramps}
\begin{align}
    \hat{\rho_i}[k+1] &= \hat{\rho_i}[k] +\frac{T}{l}( \hat{r_i}[k] - r_i[k])\label{eq:Flow-conservation_on-ramp}\\
    \check{\rho_i}[k+1] &= \check{\rho_i}[k] +\frac{T}{l}( s_i[k]-\check{s_i}[k]),\label{eq:Flow-conservation_off-ramp}
\end{align}
\end{subequations}
where $\hat{r_i}[k]$ and $\check{s_i}[k]$ are given as
\begin{subequations}\label{eq:ramp inflow outflow}
    \begin{align}
    \hat{r_i}[k]&=\min\big(w_c(\rho_m-\hat{\rho_i}[k]),v_f\rho_c,\hat{f_i}\big)\label{eq:on-ramp inflow}\\
    \check{s_i}[k]&=\min\big(v_f\check{\rho_i}[k],v_f\rho_c,\check{f_i}\big),\label{eq:off-ramp outflow}
    \end{align}
\end{subequations}
Substituting \eqref{eq:off-ramp relation}, \eqref{eq:on-ramp flow}, and \eqref{eq:CTM_flow_characteristics}  to \eqref{eq:CTM_flow_conservation} yields
\begin{align*}
\begin{split}
\hspace{-0.1cm}\rho_i[k+1] &=
\rho_i[k]+\frac{T}{l}\Big(\min\big(\delta_{i-1}[k],\sigma_{i}[k]\big)\\
&\quad-\frac{1}{\bar{\beta_i}[k]}\min\big(\delta_i[k],\sigma_{i+1}[k]\big)\\
&\quad+\min\big(v_f\hat{\rho_i}[k],\xi_i(\rho_m-\rho_i[k]),\frac{\xi_i}{w_c}v_f\rho_c\big)\Big).
\end{split}
\end{align*}
The above equation represents the traffic dynamics for highway sections connected to both on-ramp and off-ramp. Similar equations can easily be derived for other cases when highway sections are connected to single or no ramp, which are not shown here due to space constraint.
In this paper we assume that the upstream flow at section $1$ and downstream flow at section $N$ are known, denoted by $f_{\mathrm{in}}[k]$ and $f_{\mathrm{out}}[k]$ respectively. The state vector can be defined as $\m x[k] \triangleq [\rho_i[k]\hspace{1mm} \ldots\hspace{1mm}\hat{\rho_j}[k]\hspace{1mm} \ldots\hspace{1mm}\check{\rho_l}[k]\hspace{1mm} \ldots]^{\top}$ for which $i\in\Omega$, $j\in\Omega_I$ and $l\in\Omega_O$. In general, the nonlinear dynamic system can be modelled with the following equations
\begin{itemize}
\item $i\in\Omega\setminus\Omega_I\cup\Omega_O$
\begin{align*}
x_i[k+1] &=  x_i[k]+ \frac{T}{l} \Big(\min\big(\delta_{i-1}[k],\,\sigma_i[k]\big)\\
&\quad-\min\big(\delta_i[k],\,\sigma_{i+1}[k]\big)\Big)
\end{align*}

\item $i\in\Omega_I\setminus\Omega_I\cap\Omega_O$, $j=N+\bar{j}, \bar{j}\in\hat{\Omega}$
\begin{align*}
x_i[k+1] &=  x_i[k]+ \frac{T}{l} \Big(\min\big(\delta_{i-1}[k],\sigma_i[k]\big)\\
&\quad-\min\big(\delta_i[k],\,\sigma_{i+1}[k]\big)\\
&\quad+\min\big(v_fx_j[k],\xi_{i}[k](\rho_m-x_i[k]),\frac{\xi_i[k]}{w_c}v_f\rho_c\big)\Big)
\end{align*}

\item  $i\in\Omega_O\setminus\Omega_I\cap\Omega_O$, $j=N+N_I+\bar{j}, \bar{j}\in\check{\Omega}$
\begin{align*}
x_i[k+1] &=  x_i[k]+ \frac{T}{l} \Big(\min\big(\delta_{i-1}[k],\,\sigma_i[k])\big)\\
&\quad-\frac{1}{\bar{\beta}_{i}[k]}\min\big(\delta_i[k],\,\sigma_{i+1}[k]\big)\Big)
\end{align*}

\item $i\in\Omega_I\cap\Omega_O$, $j=N+\bar{j},l=N+N_I+\bar{l}, \bar{j}\in\hat{\Omega}, \bar{l}\in\check{\Omega}$
\begin{align*}
x_i[k+1] &=  x_i[k]+ \frac{T}{l} \Big(\min\big(\delta_{i-1}[k],\,\sigma_i[k]\big)\\
&\quad-\frac{1}{\bar{\beta}_{i}[k]}\min\big(\delta_i[k],\,\sigma_{i+1}[k]\big)\\
&\quad+\min\big(v_fx_j[k],\xi_{i}[k](\rho_m-x_i[k]),\frac{\xi_{i}[k]}{w_c}v_f\rho_c\big)\Big)
\end{align*}

\item $i\in\Omega_I, j=N+\bar{j},\bar{j}\in\hat{\Omega}$
\begin{align*}
x_{j}[k+1] &= x_{j}[k]+ \frac{T}{l}\big(\hat{r}_i[k] \\
&\quad-\min\big(v_fx_j[k],\xi_i[k](\rho_m-x_i[k]),\frac{\xi_i[k]}{w_c}v_f\rho_c\big)\big)
\end{align*}

\item $i\in\Omega_O,j=N+N_I+\bar{j},\bar{j}\in\check{\Omega}$
\begin{align*}
x_j[k+1] &= x_{j}[k] +\frac{T}{l}\Big(\frac{\beta_{i}[k]}{\bar{\beta}_{i}[k]}\min\big(\delta_i[k],\sigma_{i+1}[k]\big)-\check{s}_{i}[k]\Big)
\end{align*}

\end{itemize}
For $i=1$, in the above equations, we can simply replace $\delta_{i-1}[k]$ with the known inflow $f_{\mathrm{in}}[k]$, and similarly for $i=N$, we can replace $\sigma_{i+1}[k]$ with the known outflow $f_{\mathrm{out}}[k]$. The above dynamics can be expressed into a state-space, difference equation of the form
\begin{align}
&\m x[k+1] = \mA \m x[k]+\m f(\m x, \m u), \label{eq:state_space_gen}
\end{align}
\noindent where $\m u[k]=[f_{\mathrm{in}}[k]\hspace{1mm} f_{\mathrm{out}}[k]\hspace{1mm} \ldots \hspace{1mm}{\hat{f}_j[k]}\hspace{1mm} \ldots \hspace{1mm}{\check{f}_l[k]}\hspace{1mm} \ldots]^{\top}\in \mbb{R}^{m}$ for which $j\in\Omega_I$ and $l\in\Omega_O$ with $m = 2+N_I+N_O$, $\mA\in \mbb{R}^{n\times n}$  is an identity matrix of dimension $n = N+N_I+N_O$, and $\m f: \mbb{R}^{n}\times \mbb{R}^{m}\rightarrow \mbb{R}^{n}$ is a mapping that lumps all the nonlinearities, mostly comprised of min functions, in traffic density dynamics given above. 
The next section presents our strategy to perform traffic density estimation under assuming noisy measurements and model.

\section{A Robust Observer Framework}\label{sec:observer}
In the previous section, we show that highway traffic dynamics with multiple ramp flows can be modeled by a set of difference equations, which is then represented by a nonlinear state-space difference equation \eqref{eq:state_space_gen}. Since we focus on traffic density estimation, the objective of this work is to \textit{robustly} estimate the densities of all highway sections, including the ramps, under a limited number of traffic sensors or measurements. The \textit{robustness} here is defined in the sense that the estimator/observer is resilient to \textit{disturbances} which could be caused by unmodeled dynamics, unknown inputs, model uncertainty, process noise, and measurement noise---all of these exist in real-world applications. 
To that end, the traffic dynamics model with measurements and disturbance can be posed as
\begin{subequations}\label{eq:state_space_general}
	\begin{align}
	\m x[k+1] &= \mA \m x[k]+\m f(\m x, \m u)+ \m {B_{\mathrm{w}}} \m w[k] \label{eq:state_space_general_a}\\
	\m y[k] &= \mC \m x[k]+ \m {D_{\mathrm{w}}} \m w[k], \label{eq:state_space_general_b}
	\end{align}
\end{subequations}
where \eqref{eq:state_space_general_a} represents traffic dynamics with unknown inputs, \eqref{eq:state_space_general_b} is the linear measurement model with measurement noise, $\m y\in\mathbb{R}^p$ is the measurement vector, and $\m C\in\mathbb{R}^{p\times n}$ is a matrix representing the configuration and location of the traffic sensors. The disturbance vector $\m w\in\mathbb{R}^q$ lumps all unknown inputs into a single vector, with the corresponding matrices $\m {B_{\mathrm{w}}}$ and $\m {D_{\mathrm{w}}}$ are of appropriate dimensions. 

Since we utilize an observer to estimate a nonlinear dynamics \eqref{eq:state_space_general}, then it is important to study the nonlinearities appearing in \eqref{eq:state_space_general}. 
It is presumed that $\m f(\cdot)$ is locally Lipschitz continuous, which definition is given below.
\vspace{-0.1cm}
\begin{mydef}[Lipschitz Continuity]\label{def:Lipschitz_cont}
	Let $\m f : \mathbb{R}^{n}\times \mathbb{R}^{m} \rightarrow \mathbb{R}^{n}$. Then, $\m f$ is Lipschitz continuous in $(\mathbfcal{X},\mathbfcal{U})\subset \mathbb{R}^{n}\times \mathbb{R}^{m}$ if there exists a constant $\gamma \in \mathbb{R}_{+}$ such that
	\begin{align}
	\norm{\m f(\m x,\m u)-\m f(\hat{\m x},\m u)}_2 \leq \gamma \norm{\m x -  \hat{\m x} }_2, \label{eq:lipschitz}
	\end{align}
	for all $\m x, \hat{\m x} \in \mathbfcal{X}$ and $\m u \in \mathbfcal{U}$.
\end{mydef} 
\vspace{-0.1cm}
In the context of traffic dynamics model \eqref{eq:state_space_gen}, $\mathbfcal{X}$ represents the operating region of the states $\m x$. Since $\m x$ represents the densities, then $\mathbfcal{X} \triangleq [0,\rho_m]^n$. Likewise, we can deduce that $\mathbfcal{U} \triangleq [0,v_f\rho_c]^m$.
This Lipschitz constant $\gamma$ plays a role in the proposed robust observer framework, which design is discussed as follows. 

First, let $\hat{\m x}[k]$ be the estimation vector and $\hat{\m y}[k]$ be the estimation measurement vector. The observer dynamics follow a similar form to the classic Luenberger observer given as
\begin{subequations} \label{eq:nonlinear_observer_dynamics}
	\begin{align}	
	{\hat{\m x}}[k+1] &= \mA \hat{\m x}[k+1]+\m f(\hat{\m x}, \m u) + \mL(\m y[k]-\hat{\m y}[k]) \\
	\hat{\m y}[k] &= \mC \hat{\m x}[k],
	\end{align}
\end{subequations}
where $\mL(\m y-\hat{\m y})$ is the Luenberger-type correction term with $\mL\in\mathbb{R}^{n\times p}$. {\color{black} In order to ensure the existence of such observer, it is assumed that the traffic sensors have been placed in such a way that they yield the pair $(\m A,\m C)$ detectable.} By defining the estimation error as $\m e[k]  \triangleq \m x [k]- \hat{\m x}[k]$, the error dynamics can be computed as 
\begin{align}
\small \hspace{-0.3cm}{\m e} [k+1] = \left(\mA-\mL\mC\right)\m e[k] +\Delta \m f[k]+ \left(\m {B_{\mathrm{w}}}-\mL\m {D_{\mathrm{w}}}\right)\m w[k], \label{eq:est_error_dynamics}
\end{align}
where $\Delta \m f[k]\triangleq \m f(\m x, \m u)-\m f(\hat{\m x}, \m u)$.

 Our goal is to achieve asymptotic estimation error for estimation error dynamics given above. 
 In our prior work \cite{Nugroho2018TrafficJournal}, we present a robust observer using the concept of $\linf$ stability for traffic density estimation assuming nonlinear continuous-time traffic dynamics model.
The $\mathcal{L}_{\infty}$ stability theory is previously developed in \cite{pancake2002analysis} for feedback control purpose, 
which is then used in \cite{Chakrabarty2016} to develop a robust observer for nonlinear discrete-time systems which nonlinearities satisfy \textit{incremental quadratic constraint}. Here we also use the $\mathcal{L}_{\infty}$ stability concept to design a robust observer for traffic density estimation. Nonetheless in this paper, the observer is designed specifically for Lipschitz nonlinearity instead of incremental quadratic nonlinearity. Albeit incremental quadratic nonlinearity comprises a large class of nonlinear functions including Lipschitz nonlinearity, the use of condition \eqref{eq:lipschitz} will return much simpler matrix inequality formulations---for example, see \cite{Nugroho2018TrafficJournal}. To that end, we utilize the definition of $\linf$ stability presented in \cite{pancake2002analysis,Chakrabarty2016}.
It is assumed here that $\m w \in \mathcal{L}_{\infty}$ to ensure that the disturbance acting on the system is bounded, which is realistic in practical situations.
The following result presents a numerical procedure to compute the observer gain $\m L$ that, if solved,  ensures the estimation error dynamics \eqref{eq:est_error_dynamics} is $\mathcal{L}_{\infty}$ stable with performance level $\mu$. 
\vspace{-0.1cm}
\begin{theorem}\label{l_inf_theorem}
	Consider the nonlinear dynamics \eqref{eq:state_space_general} and observer \eqref{eq:nonlinear_observer_dynamics} in which $\m w \in \mathcal{L}_{\infty}$ and $\m f : \mathbb{R}^{n}\times \mathbb{R}^{m} \rightarrow \mathbb{R}^{n}$ is locally Lipschitz in $(\mathbfcal{X},\mathbfcal{U})$ with Lipschitz constant $\gamma$. If there exist $\mP\in\mathbb{S}^n_{++}$, $\mY\in\mathbb{R}^{n\times p}$, $\epsilon,\mu_0,\mu_1,\mu_2\in \mathbb{R}_{+}$, and $\alpha \in \mathbb{R}_{++}$ so that the following optimization problem is solved
		\begin{subequations}\label{eq:l_inf_theorem2}
			\vspace{-0.0cm}
			\begin{align}
			&\minimize_{\m P, \m Y, \epsilon, \alpha, \mu_{0,1,2}} \quad \mu_0\mu_1 + \mu_2 \label{eq:l_inf_theorem2_0}\\
			&\subjectto \;\;\nonumber \\
			&\bmat{  (\alpha-1)\mP+\epsilon\gamma^2\mI&*&*&*\\
				\mP\mA-\mY\mC& \mP -\epsilon\mI &*&*\\
				\mO & \m\Phi^{\top} & -\alpha\mu_0\mI &*\\
				\mP\mA-\mY\mC&\mO& \m\Phi &-\mP} \preceq 0 \label{eq:l_inf_theorem2_1}\\
			&\bmat{-\mP & * & * \\
				\mO & -\mu_2\mI & *\\
				\mZ & \mO & -\mu_1\mI}\preceq 0, \label{eq:l_inf_theorem2_2}
			\end{align}
		\end{subequations}
\noindent where $\m\Phi \triangleq \mP\m {B_{\mathrm{w}}}-\mY\m {D_{\mathrm{w}}}$, 
then the error dynamics given in \eqref{eq:est_error_dynamics} is $\mathcal{L}_{\infty}$ stable with performance level $\mu = \sqrt{\mu_0\mu_1+\mu_2}$ for performance output given as $\m z = \mZ \m e$, where $\mZ\in\mathbb{R}^{z\times n}$, and observer gain $\mL$ computed as $\mL = \mP^{-1}\mY$.
\end{theorem}   
\vspace{-0.1cm}
\vspace{-0.1cm}
We now note that the optimization problem described in \eqref{eq:l_inf_theorem2} is nonconvex due to bilinearity appearing in the form of $(\alpha-1)\mP$, $\alpha\mu_0$, and $\mu_0\mu_1$. To get a convex one, we simply set $\alpha$ and $\mu_0$ or $\mu_1$ a priori. and solve for the other variables using any semidefinite programming (SDP) solver.  

\begin{figure}[t]
	\vspace{-0.0cm}
	\centering 
	{\includegraphics[keepaspectratio=true,scale=0.46]{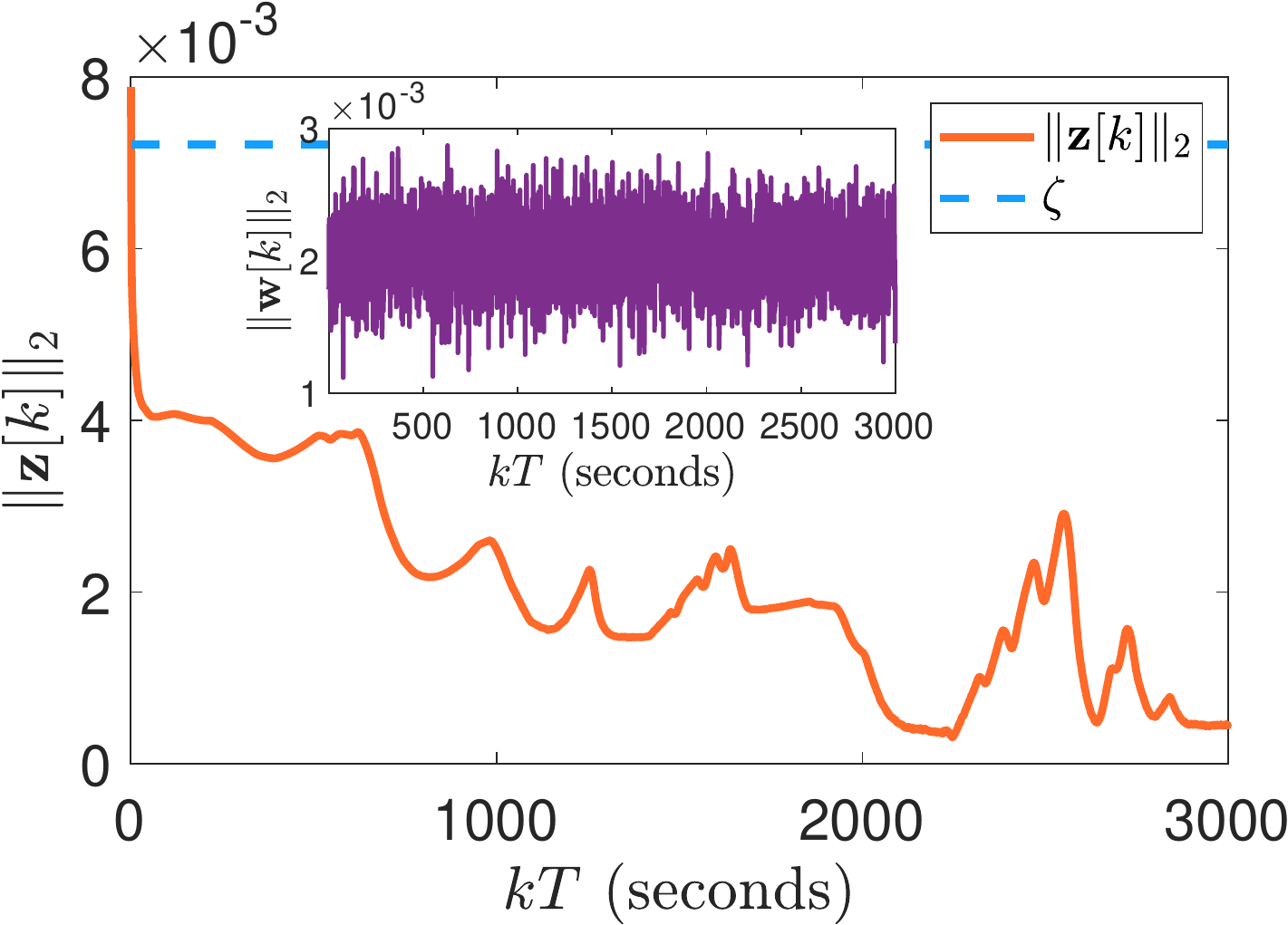}}\vspace{-0.2cm}
	\caption{Comparison between the norm of performance output $\Vert{\m z}(t)\Vert_2$ and disturbance where $\zeta \triangleq \mu\Vert{\m w}(t)\Vert_{\mathcal{L}_{\infty}}$.}
	\label{fig:performance_case_10}\vspace{-0.5cm}
\end{figure}

\begin{figure*}
	\vspace{-0.0cm}
	\centering 
	\subfloat[]{\includegraphics[keepaspectratio=true,scale=0.478]{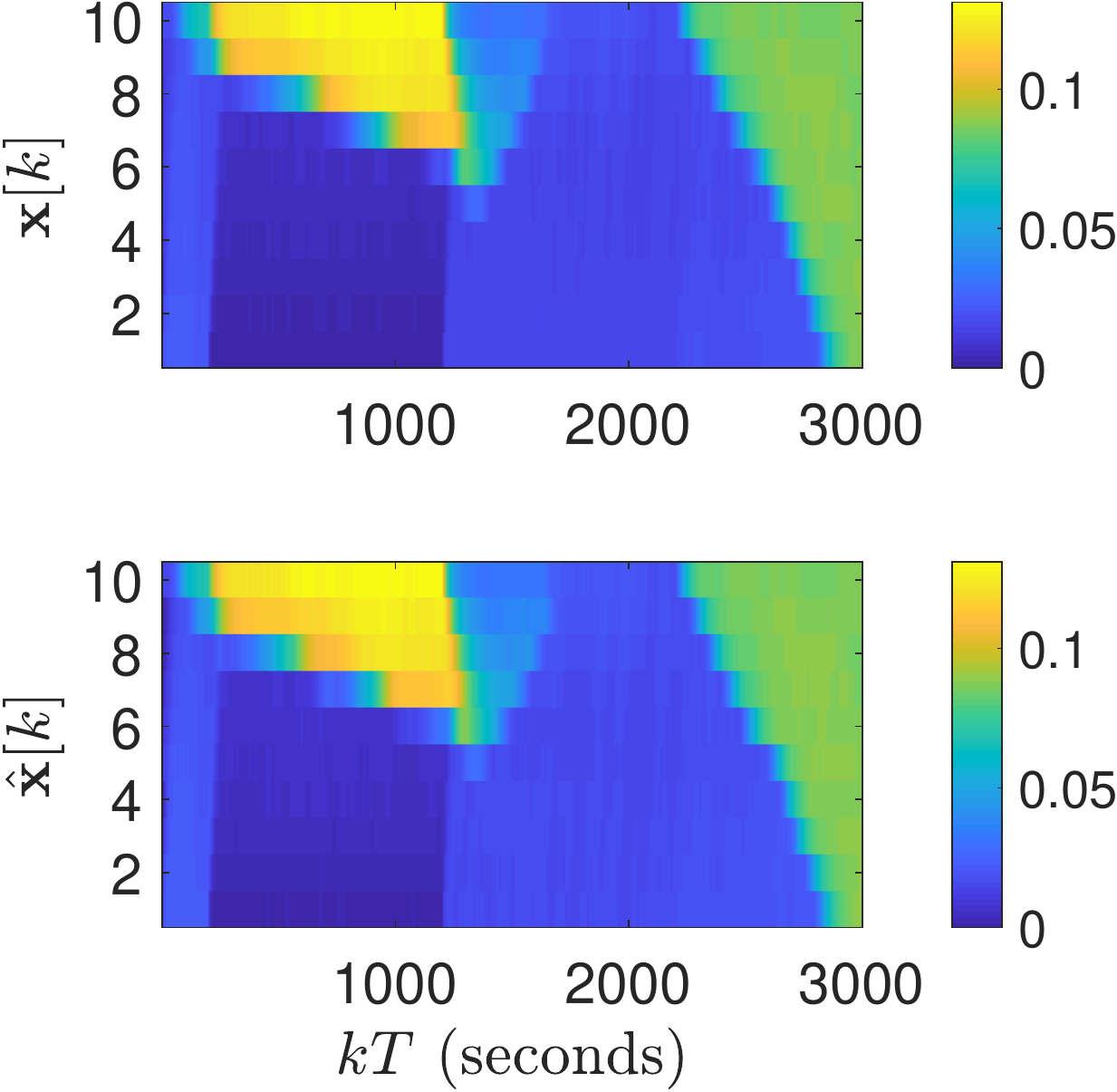}}{}\vspace{-0.0cm}
	\subfloat[]{\includegraphics[keepaspectratio=true,scale=0.478]{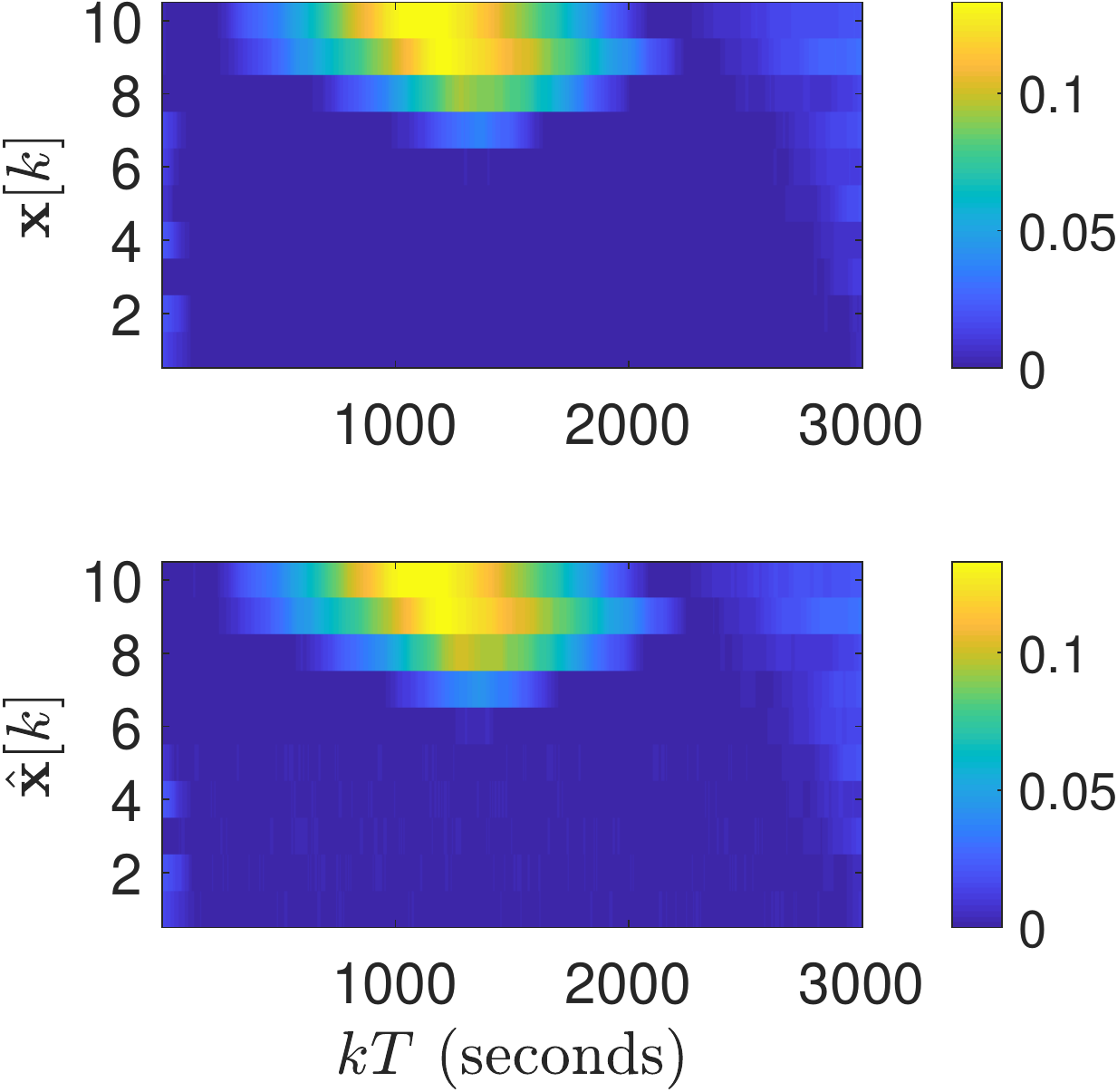}}{}\hspace{-0.0cm}
	\subfloat[]{\includegraphics[keepaspectratio=true,scale=0.478]{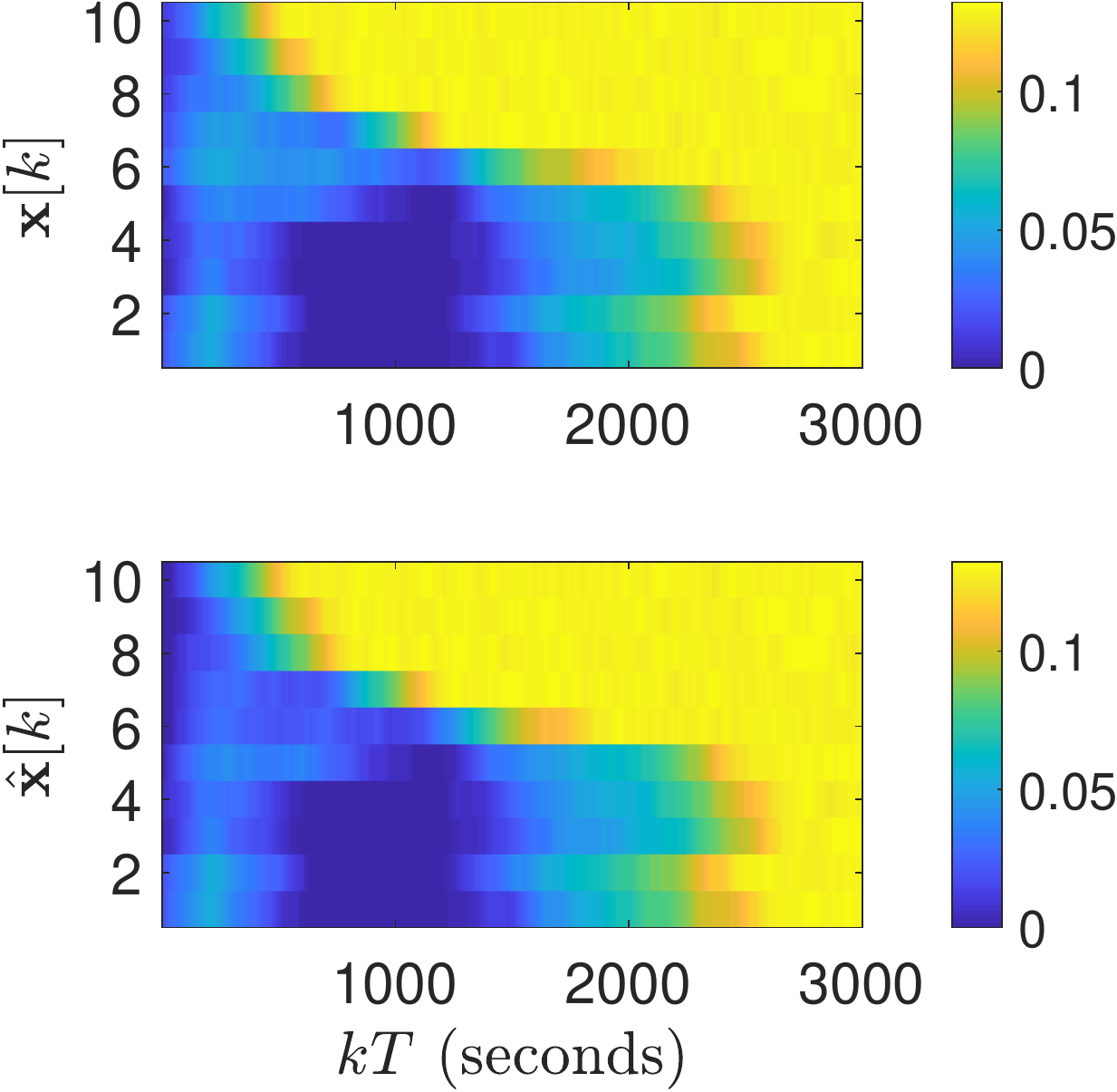}}{}\hspace{-0.0cm}\vspace{-0.1cm}
	\caption{Comparison between real ($\m x [k]$) and estimated ($\hat{\m x} [k]$) traffic densities on (a) highway sections, (b) on-ramps, and (c) off-ramps.}
	\label{fig:density_evolution_case_10}\vspace{-0.1cm}
\end{figure*} 

\section{Numerical Study}\label{sec:numerical}
\subsection{Evaluation of $\linf$ Observer for Traffic Density Estimation}\label{sec:test_linf_obs}
This section demonstrates the effectiveness of the proposed $\linf$ observer framework to estimate traffic density under disturbances in the measurement model. All simulations are carried out using MATLAB R2019a running on a 64-bit Windows 10 with 3.4GHz Intel\textsuperscript{R} Core\textsuperscript{TM} i7-6700 CPU and 16 GB of RAM with YALMIP \cite{Lofberg2004} as the interface to solve all convex SDPs. In this numerical test we consider a simple highway consisting characterized by the following quantities.
\begin{itemize}
	\item Network parameters: $v_f = 28.8889$ m/s ($65$ mph), $w_c = 6.6667$ m/s ($10$ mph), $\rho_c = 0.0249$ vehicles/m ($40$ vehicles/mile), $\rho_m = 0.1333$ vehicles/m, $L = 2000$ m, and $l = 200$ m. With time step $T = 1$ s, the CFL condition is found to be satisfied. 
	\item A total of $n=30$ states, with $N=10$ highway sections where each section is connected to both on- and off-ramps. 
	\item There are $p = 13$ number of traffic sensors installed in total and distributed on the highway sections, on-ramps, and off-ramps, where $3$ sensors on the $\nth{2}$, $\nth{5}$, $\nth{10}$ section, $5$ sensors on the $\nth{2}$, $\nth{4}$, $\nth{5}$, $\nth{7}$, $\nth{9}$ on-ramps, and $5$ sensors on the $\nth{1}$, $\nth{3}$, $\nth{6}$, $\nth{8}$, $\nth{10}$ off-ramps such that $\m C \in \mathbb{R}^{13 \times 30}$. Notice that the highway is \textit{under-sensed}, in the sense that less than half of the total highway sections are equipped with traffic sensors. We also note that to ensure robustness against uncertainty, it is conceivable to require more sense measurements, thereby hedging uncertainty.
	\item The upstream, downstream, on-ramps inflows, and out-ramp outflows are constructed in a particular way to make it varying and random throughout the entire simulation window. The spit ratio for out-ramps is set to be $10\%$.
\end{itemize}
We aim to estimate the traffic density on all highway sections that are not equipped with traffic sensors. The corresponding Lipschitz constant for this case is chosen to be $\gamma = 0.5$. To obtain a convex problem, we set $\alpha = 0.05$ and $\mu_1 = 10^4$. We use SDPNAL+ \cite{Yang2015} to solve the resulting convex problem. The performance matrix is chosen to be $\mZ = 0.1\mI$. 
Herein, we consider the case when there exists Gaussian noise on the measurements having covariance matrix equal to $R = r\mI$ with $r = 10^{-3}$. 

After successfully solving optimization problem described in Theorem \ref{l_inf_theorem}, we obtain the observer gain matrix $\mL$ with the corresponding performance index $\mu = 2.5092$. The  computed $\linf$ norm of the disturbance for this case is $\norm{\m w}_{\mathcal{L}_{\infty}}=2.875\times 10^{-3}$. The upper bound on the norm of performance vector $\m z$ can then be computed as $\zeta \triangleq \mu\norm{\m w}_{\mathcal{L}_{\infty}} = 7.214\times 10^{-3}$. From Fig. \ref{fig:performance_case_10} it can be seen that the $\linf$ stability for the estimation error dynamics \eqref{eq:est_error_dynamics} is satisfied, that is, $\norm{\m z}_2$ converges to a value below $\zeta$. Fig. \ref{fig:density_evolution_case_10} illustrates the evolution of the real and estimated density for all highway sections, including the ramps. This empirically shows that the proposed robust $\linf$ observer framework is able to perform traffic density estimation for the given model with a satisfactory performance. 

\begin{figure}[t]
	\vspace{-0.0cm}
	\centering 
	{\includegraphics[keepaspectratio=true,scale=0.46]{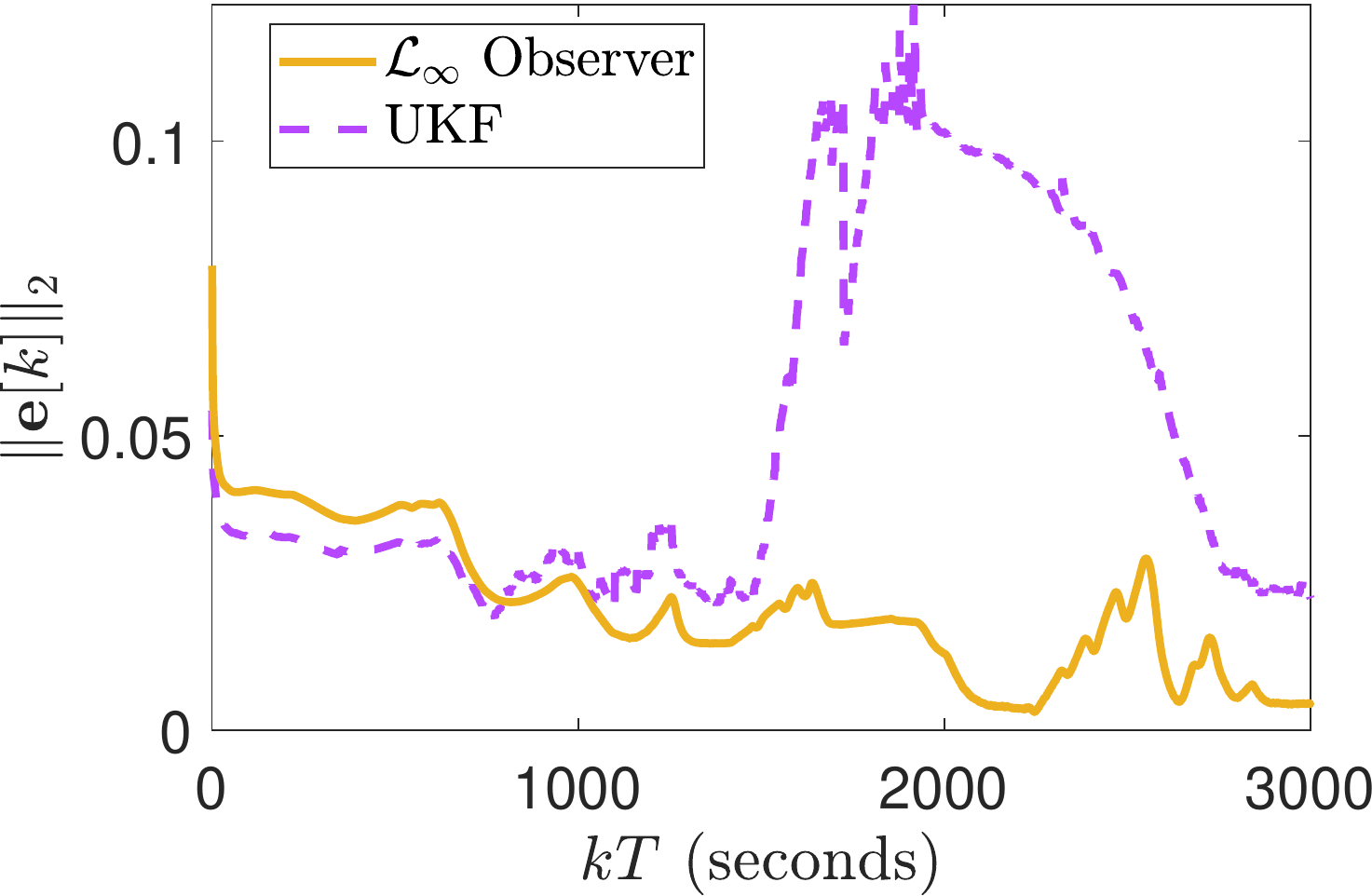}}\vspace{-0.25cm}
	\caption{Trajectories of estimation error norm for $\linf$ observer and UKF.}
	\label{fig:ukf_comparison}\vspace{-0.25cm}
\end{figure}

\subsection{Comparison with Unscented Kalman Filter (UKF)}\label{sec:ukf_comparison} 
In this section we compare the $\linf$ observer with Unscented Kalman Filter (UKF) in estimating traffic density. UKF is considered here because \textit{(a)} UKF has been widely used for traffic density estimation in the literature, e.g. \cite{Hegyi2006}, and \textit{(b)} unlike Extended Kalman Filter (EKF) that relies on the linearization and Jacobian matrix of the nonlinear dynamics, UKF is a derivative free estimator which is suitable for the non-continuously differentiable traffic dynamics model developed in this paper. The process and measurement noise covariance matrices for UKF are determined to be $Q = q\mI$ with $q = 10^{-3}$ and $R = r\mI$ with $r = 10^{-3}$, in which the initial error covariance is set to be $\m P_{cov,0} = 10^{-4}\mI$. The constants to determine sigma points for UKF, which is needed in unscented transformation \cite{wan2000unscented}, are set to be $\alpha = 0.01$, $\beta = 2$, and $\kappa = -4$. 

\begin{table}[t]
	\footnotesize	\centering \renewcommand{\arraystretch}{1.4}
		\begin{threeparttable}[b]
		\caption{RMSE and simulation running time for $\linf$ Observer and UKF.}\vspace{-0.2cm}
	\begin{tabular}{|l|c|c|}
		\hline
		\textbf{State Estimator}	& \textbf{RMSE}  & \textbf{Simulation Running Time}  \\ \hline \hline
		$\linf$ Observer\tnote{$\dagger$}	& $0.0868$ & $3.422$ seconds \\ \hline
		UKF	& $0.1589$ & $70.635$ seconds \\ \hline
	\end{tabular}\label{tab:ukf_comparison}
	\begin{tablenotes}
	\item[$\dagger$] It took $0.0993$ seconds to obtain the observer gain matrix. 
\end{tablenotes} 
\end{threeparttable}
	\vspace{-0.4cm}
\end{table}

The result of this comparison is depicted in Fig. \ref{fig:ukf_comparison}, which shows the corresponding error norm for $\linf$ observer and UKF. We see that the performance of $\linf$ observer and UKF are very similar, albeit UKF exhibits a slightly larger estimation error. To obtain quantitative comparisons, we numerically measure their respective \textit{Root Mean Square Error} (RMSE), defined in the following formulation 
\begin{align*}
\mathrm{RMSE} &= \sum_{i=1}^{n} \sqrt{\frac{1}{k_f}\sum_{k=1}^{k_f}(e_i[k])^2},
\end{align*} 
where $k_f = 3000$, and their simulation running time, which results are summarized in Tab. \ref{tab:ukf_comparison}. From this table we see that UKF yields a slightly larger RMSE compared to $\linf$ observer. This corroborates the smaller $\linf$ observer's estimation error trajectory illustrated in Fig. \ref{fig:ukf_comparison}.
Despite of this, our numerical test results indicate that $\linf$ observer actually shows a much faster simulation running time. This is due to the fact that in observer framework, a constant estimator gain matrix is used, which is in contrast with UKF as its estimator gain matrix has to be computed in every iteration. This renders the $\linf$ observer to be computationally attractive and in turn, makes it much more suitable for practical implementation compared to UKF for large-scale simulations. Moreover, unlike UKF, $\linf$ observer provides a theoretical guarantee on the deviation of the performance vector $\m z$ (which is nothing but a linear combination of the estimation error $\m e$) with respect to disturbance $\m w$. This can be beneficial for the system operator who wants a certain performance guarantee on how far off are the state estimates from actual states. 


\section{Paper's Limitations and Future Work}\label{sec:summary}

This paper presents a control-theoretic approach to perform traffic density estimation on ramp-connected, stretched highways using the asymmetric cell transmission model under process and measurement uncertainty. The estimator dynamics are akin to the vintage Luenberger observer, making the presented estimator amenable to real-time simulations for large-scale traffic networks. 

The paper's limitation are two-fold. First, parametric uncertainty in the observer design and SDP formulation are not considered and second, it is certainty conceivable to further optimize the performance level $\mu$ rather than fixing some constants by solving the nonconvex problem~\eqref{eq:l_inf_theorem2}. That is, tight convex approximations for~\eqref{eq:l_inf_theorem2} can return an improved performance level---and hence an improved state estimation. 

\section{Acknowledgments}

We gratefully acknowledge the constructive feedback and thorough suggestions from the four reviewers and the editor. This work was partially supported by the National Science Foundation under Grants 1636154, 1728629, 1917164, and 1917056


\bibliographystyle{IEEEtran}	\bibliography{bibl}

\end{document}